\documentclass{ws-ijmpe}
\usepackage{tipa}
\begin{document}
\markboth{Jameel-Un Nabi}{Neutrino energy loss rates due to titanium
isotopes}
\catchline{}{}{}{}{}
\title{Neutrino and antineutrino energy loss rates in massive stars
due to isotopes of titanium}
\author{\footnotesize Jameel-Un Nabi}
\address{Faculty of Engineering Sciences, GIK Institute of Engineering
Sciences and Technology, Topi 23640, NWFP, Pakistan\\
jnabi00@gmail.com}
 \maketitle
\begin{history}
\received{(received date)}
\revised{(revised date)}
\end{history}
\begin{abstract}
Weak interaction rates on titanium isotopes are important during the
late phases of evolution of massive stars. A search was made for key
titanium isotopes from available literature and a microscopic
calculation of weak rates of these nuclei were performed using the
proton-neutron quasiparticle random phase approximation (pn-QRPA)
theory. Earlier the author presented the stellar electron capture
rates on titanium isotopes. In this paper I present the neutrino and
antineutrino energy loss rates due to capture and decay rates on isotopes of
titanium in stellar environment. Accurate estimate of neutrino
energy loss rates are needed for the study of the late stages of the
stellar evolution, in particular for cooling of neutron stars and
white dwarfs. The results are also compared against previous
calculations. At high stellar temperatures the calculated neutrino
and antineutrino energy loss rates are bigger by more than two
orders of magnitude as compared to the large scale shell model
results and favor stellar cores with lower entropies. This study can
prove useful for core-collapse simulators.
\end{abstract}

\section{Introduction}
The classical work on energy transport by neutrinos and
antineutrinos in non-rotating massive stars was performed by Colgate
$\&$ White \cite{Col66} and Arnett \cite{Arn67}. Today, despite
considerable advancement in the available technology, the collapse
simulators find it challenging to generate an explosion out of the
collapsing core of massive stars. The prompt shock that follows the
bounce of the core stagnates and is not possible to cause a
supernova explosion on its own. It loses energy in disintegrating
iron nuclei and through neutrino emissions (mainly non-thermal)
which are till then transparent to the stellar matter. Various
energizing mechanisms for shock revival have been proposed in the
text. These include, but are not limited to, the "preheating"
mechanism proposed by Haxton \cite{Hax88}, inclusion of magnetic
fields (e.g. Ref. \cite{Kot04}) and rotations (e.g. Ref.
\cite{Wal05}) in the simulation codes. However to date there have
been no successfully simulated spherically symmetric explosions.
Even the 2D simulations (addition of convection) performed with a
Boltzmann solver for the neutrino transport fails to convert the
collapse into an explosion \cite{Bur03}.

Neutrinos radiate around 10$\%$ of the rest mass converting the star
to a neutron star. Initially the nascent neutron star is a hot
thermal bath of dense nuclear matter, $e^{-}e^{+}$ pairs, photons
and neutrinos. Neutrinos, having the weak interaction, are most
effective in cooling and diffuse outward within a few seconds, and
eventually escape with about 99$\%$ of the released gravitational
energy. Despite the small neutrino-nucleus cross sections, the
neutrinos flux generated by the cooling of a neutron star can
produce a number of nuclear transmutations as it passes the
onion-like structured envelope surrounding the neutron star.
Neutrinos from core-collapse supernovae are unique messengers of the
microphysics of supernovae. They provide information regarding the
neutronization due to electron capture, the infall phase, the
formation and propagation of the shock wave and the cooling phase.
Cooling rate is one of the crucial parameters that strongly affect
the stellar evolution. During the late stages of stellar evolution a
star mainly looses energy through neutrinos. White dwarfs and
supernovae (which are the endpoints for stars of varying masses)
have both cooling rates largely dominated by neutrino production. A
cooling proto-neutron star emits about $3 \times 10^{53}$ erg in
neutrinos, with the energy roughly equipartitioned among all three
species. The neutrinos and antineutrinos produced as a result of
nuclear reactions are transparent to the stellar matter at
presupernova densities and therefore assist in cooling the core to a
lower entropy state. This scenario does not necessarily hold at
extremely high densities and temperatures (this would be the case
for stellar collapse where dynamical time scales become shorter than
the neutrino transport time scales) where neutrinos can become
trapped in the so-called neutrinospheres mainly due to elastic
scattering with nuclei. Prior to stellar collapse one requires an
accurate determination of neutrino energy loss rates in order to
perform a careful study of the final branches of star evolutionary
tracks.

The neutrino and antineutrino energy loss rates can occur through
four different weak-interaction mediated channels: electron and
positron emissions, and, continuum electron and positron captures.
The stellar neutrinos are produced due to electron captures and
positron decays whereas the antineutrinos are produced due to beta
decays and positron captures. As discussed above these neutrinos
(antineutrinos) are transparent at presupernova densities and escape
the site and help the core achieve a lower entropy which is of vital
importance for the core to explode later.

Nabi and Klapdor-Kleingrothaus \cite{Nab99} first reported the
calculation of weak interaction rates for 709 nuclei with A = 18 to
100 in stellar environment using the pn-QRPA theory. The authors
then presented a detailed calculation of stellar weak interaction
rates over a wide range of temperature and density scale for
fp/fpg-shell nuclei \cite{Nab04}. Because of the high temperatures
prevailing during the presupernova and supernova phase of a massive
star, there is a reasonable probability of occupation of parent
excited states and the total weak interaction rates have a finite
contribution form these excited states. The pn-QRPA theory allows a
microscopic state-by-state calculation of \textit{all} these partial
rates and this feature of the model greatly enhances the reliability
of the calculated rates in stellar matter. The pn-QRPA model can
handle any arbitrarily heavy system of nucleons as it has access to
a luxurious model space of up to $7\hbar\omega$ shells. The pn-QRPA
model was successfully used to calculate weak interaction rates on
important iron-regime nuclei (e.g. Refs.
\cite{Nab05,Nab07a,Nab08,Nab09}).

Weak interaction rates on hundreds of nuclei are involved in the
complex dynamics of stellar evolution culminating in a supernova
explosion. A search for key weak interaction nuclei in presupernova
evolution was performed by Aufderheide and collaborators
\cite{Auf94}. Late phases of evolution (namely after core silicon
burning) in massive stars were considered and a search was performed
for the most important electron captures and $\beta$-decay nuclei in
these scenarios. The lists consisted of dozens of iron-regime
nuclei. From these lists electron captures on $^{49,51,52,53,54}$Ti
and $\beta$-decay of $^{51,52,53,54,55,56}$Ti were short-listed to
be of astrophysical importance. Previously Nabi and collaborators
\cite{Nab07} presented a detailed analysis of the calculation of
stellar electron capture rates on twenty two titanium isotopes. Out
of these twenty two isotopes of titanium seven isotopes, namely
$^{49,51,52,53,54,55,56}$Ti, are suggested to be important in
cooling the core of massive stars through the (anti)neutrino
produced via the weak interaction reactions. In this paper I present
the neutrino and antineutrino energy loss rates due to these seven
isotopes of titanium in stellar matter. The next section discusses
briefly the formalism and presents the calculated neutrino and
antineutrino energy loss rates. Comparison with previous
calculations is also presented in this section. Section 3 finally
summarizes the main conclusions.

\section{Calculations and Results}
The Hamiltonian of the pn-QRPA model and its diagonalization was
discussed earlier in Ref. \cite{Nab07}. As mentioned in the previous
section the neutrino and antineutrino energy loss rates can occur
through four different weak-interaction mediated channels: electron
and positron emissions, and, continuum electron and positron
captures. It is assumed that the neutrinos and antineutrinos
produced as a result of these reactions are transparent to the
stellar matter during the presupernova evolutionary phases and
contributes effectively in cooling the system. The neutrino and
antineutrino energy loss rates were calculated using the relation

\begin{equation}
\lambda ^{^{\nu(\bar{\nu})}} _{ij} = \left(\frac{ln 2}{D} \right)
[f_{ij}^{\nu} (T, \rho, E_{f})][B(F)_{ij} + (g_{A}/ g_{V})^{2}
B(GT)_{ij}]. \label{wi}
\end{equation}
The value of D was taken to be 6295s \cite{Yos88}. $B_{ij}'s$ are
the sum of reduced transition probabilities of the Fermi B(F) and
Gamow-Teller (GT) transitions B(GT). The effective ratio of axial
and vector coupling constants, $(g_{A}/g_{V})$ was taken to be
-1.254 \cite{Rod06}. The $f_{ij}^{\nu}$ are the phase space
integrals and are functions of stellar temperature ($T$), density
($\rho$) and Fermi energy ($E_{f}$) of the electrons. They are
explicitly given by
\begin{equation}
f_{ij}^{\nu} \, =\, \int _{1 }^{w_{m}}w\sqrt{w^{2} -1} (w_{m} \,
 -\, w)^{3} F(\pm Z,w)(1- G_{\mp}) dw,
\label{phdecay}
\end{equation}
and by
\begin{equation}
f_{ij}^{\nu} \, =\, \int _{w_{l} }^{\infty }w\sqrt{w^{2} -1} (w_{m}
\,
 +\, w)^{3} F(\pm Z,w)G_{\mp} dw.
\label{phcapture}
\end{equation}
In Eqs. ~(\ref{phdecay}) and ~(\ref{phcapture})  $w$ is the total
energy of the electron including its rest mass, $w_{l}$ is the total
capture threshold energy (rest+kinetic) for positron (or electron)
capture. F($ \pm$ Z,w) are the Fermi functions and were calculated
according to the procedure adopted by Gove and Martin \cite{Gov71}.
G$_{\pm}$ is the Fermi-Dirac distribution function for positrons
(electrons).
\begin{equation}
G_{+} =\left[\exp \left(\frac{E+2+E_{f} }{kT}\right)+1\right]^{-1},
\label{Gp}
\end{equation}
\begin{equation}
 G_{-} =\left[\exp \left(\frac{E-E_{f} }{kT}
 \right)+1\right]^{-1},
\label{Gm}
\end{equation}
here $E$ is the kinetic energy of the electrons and $k$ is the
Boltzmann constant.

For the decay (capture) channel Eq. ~(\ref{phdecay}) (Eq.
~(\ref{phcapture})) was used for the calculation of phase space
integrals. Upper (lower) signs were used for the case of electron
(positron) emissions in Eq.~(\ref{phdecay}). Similarly upper (lower)
signs were used for the case of continuum electron (positron)
captures in Eq. ~(\ref{phcapture}). Details of the calculation of
reduced transition probabilities can be found in Ref. \cite{Nab04}.
Construction of parent and daughter excited states and calculation
of transition amplitudes between these states can be seen in Ref.
\cite{Nab99a}.

The total neutrino energy loss rate per unit time per nucleus is
given by
\begin{equation}
\lambda^{\nu} =\sum _{ij}P_{i} \lambda _{ij}^{\nu}, \label{nurate}
\end{equation}
where $\lambda_{ij}^{\nu}$ is the sum of the electron capture and
positron decay rates for the transition $i \rightarrow j$ and
$P_{i}$ is the probability of occupation of parent excited states
which follows the normal Boltzmann distribution.

On the other hand the total antineutrino energy loss rate per unit
time per nucleus is given by
\begin{equation}
\lambda^{\bar{\nu}} =\sum _{ij}P_{i} \lambda _{ij}^{\bar{\nu}},
\label{nubarrate}
\end{equation}
where $\lambda_{ij}^{\bar{\nu}}$ is the sum of the positron capture
and electron decay rates for the transition $i \rightarrow j$.

\begin{table}[pt]
\tbl{Neutrino and antineutrino energy loss rates due to
$^{49,51,52,53}$Ti for selected densities and temperatures in
stellar matter. log$\rho Y_{e}$ has units of $g/cm^{3}$, where
$\rho$ is the baryon density and $Y_{e}$ is the ratio of the lepton
number to the baryon number. Temperatures ($T_{9}$) are given in
units of $10^{9}$ K. $\lambda_{\nu}$ ($\lambda_{\bar{\nu}}$) are the
total neutrino (antineutrino) energy loss rates  as a result of
$\beta^{+}$ decay and electron capture ($\beta^{-}$ decay and
positron capture) in units of $MeV s^{-1}$. All calculated rates are
tabulated in logarithmic (to base 10) scale. In the table, -100
means that the rate is smaller than 10$^{-100} MeV s^{-1}$.}
{\scriptsize\begin{tabular}{|cc|cccccccc|} $log\rho Y_{e}$ & $T_{9}$
& \multicolumn{2}{|c|}{$^{49}$Ti}& \multicolumn{2}{|c|}{$^{51}$Ti} &
\multicolumn{2}{|c|}{$^{52}$Ti}
& \multicolumn{2}{|c|}{$^{53}$Ti} \\
\cline{3-10} & &  \multicolumn{1}{c}{$\lambda_{\nu}$} &
\multicolumn{1}{|c|}{$\lambda_{\bar{\nu}}$} &
\multicolumn{1}{c}{$\lambda_{\nu}$} &
\multicolumn{1}{|c|}{$\lambda_{\bar{\nu}}$} &
\multicolumn{1}{c}{$\lambda_{\nu}$} &
\multicolumn{1}{|c|}{$\lambda_{\bar{\nu}}$} &
\multicolumn{1}{c}{$\lambda_{\nu}$} &
\multicolumn{1}{|c|}{$\lambda_{\bar{\nu}}$} \\ \hline

 1.0  &   0.01  & -100  & -100  &  -100  &   -2.607  & -100  &   -3.449  & -100  &   -1.804  \\
 1.0  &   0.10  & -100  &  -59.070  &  -100  &   -2.637  & -100  &   -3.449  & -100  &   -1.804  \\
 1.0  &   0.20  &  -61.794  &  -32.015  &  -100  &   -2.677  & -100  &   -3.449  & -100  &   -1.802  \\
 1.0  &   0.40  &  -35.769  &  -17.468  &   -91.759  &   -2.708  & -100  &   -3.449  & -100  &   -1.763  \\
 1.0  &   0.70  &  -23.720  &  -10.992  &   -55.415  &   -2.723  &  -73.852  &   -3.449  &  -71.037  &   -1.691  \\
 1.0  &   1.00  &  -17.711  &   -9.082  &   -39.656  &   -2.724  &  -52.450  &   -3.448  &  -50.164  &   -1.645  \\
 1.0  &   1.50  &  -12.765  &   -7.354  &   -27.028  &   -2.686  &  -35.482  &   -3.431  &  -33.669  &   -1.604  \\
 1.0  &   2.00  &  -10.130  &   -6.295  &   -20.455  &   -2.596  &  -26.795  &   -3.337  &  -25.268  &   -1.575  \\
 1.0  &   3.00  &   -7.258  &   -4.942  &   -13.560  &   -2.356  &  -17.768  &   -2.791  &  -16.639  &   -1.336  \\
 1.0  &   5.00  &   -4.550  &   -3.355  &    -7.634  &   -1.667  &   -9.957  &   -1.597  &   -9.321  &   -0.320  \\
 1.0  &  10.00  &   -1.609  &   -1.099  &    -2.483  &   -0.095  &   -3.249  &    0.239  &   -2.939  &    0.904  \\
 1.0  &  30.00  &    2.887  &    2.888  &     3.065  &    3.660  &    2.699  &    3.560  &    3.149  &    4.068  \\
 4.0  &   0.01  & -100  & -100  &  -100  &   -2.608  & -100  &   -3.452  & -100  &   -1.805  \\
 4.0  &   0.10  & -100  &  -62.245  &  -100  &   -2.638  & -100  &   -3.451  & -100  &   -1.805  \\
 4.0  &   0.20  &  -58.738  &  -35.071  &  -100  &   -2.678  & -100  &   -3.451  & -100  &   -1.802  \\
 4.0  &   0.40  &  -32.754  &  -20.483  &   -88.744  &   -2.708  & -100  &   -3.450  & -100  &   -1.764  \\
 4.0  &   0.70  &  -21.352  &  -13.359  &   -53.047  &   -2.723  &  -71.484  &   -3.450  &  -68.669  &   -1.691  \\
 4.0  &   1.00  &  -16.661  &  -10.131  &   -38.606  &   -2.725  &  -51.400  &   -3.450  &  -49.114  &   -1.645  \\
 4.0  &   1.50  &  -12.607  &   -7.511  &   -26.870  &   -2.687  &  -35.324  &   -3.437  &  -33.511  &   -1.605  \\
 4.0  &   2.00  &  -10.094  &   -6.329  &   -20.420  &   -2.597  &  -26.759  &   -3.345  &  -25.232  &   -1.575  \\
 4.0  &   3.00  &   -7.251  &   -4.947  &   -13.554  &   -2.356  &  -17.762  &   -2.795  &  -16.633  &   -1.336  \\
 4.0  &   5.00  &   -4.548  &   -3.355  &    -7.632  &   -1.668  &   -9.956  &   -1.598  &   -9.319  &   -0.320  \\
 4.0  &  10.00  &   -1.608  &   -1.098  &    -2.482  &   -0.094  &   -3.247  &    0.239  &   -2.938  &    0.904  \\
 4.0  &  30.00  &    2.889  &    2.890  &     3.067  &    3.662  &    2.700  &    3.562  &    3.151  &    4.070  \\
 7.0  &   0.01  & -100  & -100  &  -100  &   -2.939  & -100  &   -3.907  & -100  &   -1.895  \\
 7.0  &   0.10  &  -74.334  &  -87.805  &  -100  &   -2.960  & -100  &   -3.906  & -100  &   -1.895  \\
 7.0  &   0.20  &  -40.507  &  -46.899  &  -100  &   -2.990  & -100 &   -3.905  & -100  &   -1.893  \\
 7.0  &   0.40  &  -22.993  &  -26.402  &   -78.982  &   -3.011  & -100  &   -3.900  & -100  &   -1.850  \\
 7.0  &   0.70  &  -15.052  &  -17.539  &   -46.746  &   -3.016  &  -65.184  &   -3.888  &  -62.368  &   -1.770  \\
 7.0  &   1.00  &  -11.663  &  -13.895  &   -33.608  &   -3.006  &  -46.402  &   -3.871  &  -44.116  &   -1.719  \\
 7.0  &   1.50  &   -8.825  &  -10.799  &   -23.089  &   -2.936  &  -31.543  &   -3.833  &  -29.729  &   -1.673  \\
 7.0  &   2.00  &   -7.273  &   -8.936  &   -17.598  &   -2.805  &  -23.939  &   -3.792  &  -22.411  &   -1.640  \\
 7.0  &   3.00  &   -5.541  &   -6.528  &   -11.844  &   -2.504  &  -16.052  &   -3.665  &  -14.923  &   -1.391  \\
 7.0  &   5.00  &   -3.846  &   -3.945  &    -6.929  &   -1.764  &   -9.253  &   -2.224  &   -8.616  &   -0.360  \\
 7.0  &  10.00  &   -1.510  &   -1.189  &    -2.383  &   -0.161  &   -3.149  &    0.167  &   -2.839  &    0.867  \\
 7.0  &  30.00  &    2.892  &    2.887  &     3.070  &    3.659  &    2.704  &    3.559  &    3.155  &    4.067  \\
10.0  &   0.01  &    2.448  & -100  &     0.705  & -100  &   -2.099  & -100  &   -2.444  & -100  \\
10.0  &   0.10  &    2.451  & -100  &     0.646  & -100  &   -2.093  & -100  &   -2.440  & -100  \\
10.0  &   0.20  &    2.452  & -100  &     0.569  & -100  &   -2.092  & -100  &   -2.439  & -100  \\
10.0  &   0.40  &    2.459  & -100  &     0.503  & -100  &   -2.087  & -100  &   -2.431  &  -75.486  \\
10.0  &   0.70  &    2.481  &  -85.387  &     0.474  &  -63.156  &   -2.074  &  -66.149  &   -2.412  &  -44.389  \\
10.0  &   1.00  &    2.504  &  -60.612  &     0.540  &  -44.980  &   -2.051  &  -46.814  &   -2.393  &  -31.731  \\
10.0  &   1.50  &    2.540  &  -41.152  &     0.910  &  -30.618  &   -1.973  &  -31.527  &   -2.359  &  -21.662  \\
10.0  &   2.00  &    2.573  &  -31.298  &     1.281  &  -23.292  &   -1.827  &  -23.724  &   -2.284  &  -16.478  \\
10.0  &   3.00  &    2.626  &  -21.274  &     1.733  &  -15.781  &   -1.364  &  -15.715  &   -1.280  &  -11.090  \\
10.0  &   5.00  &    2.709  &  -13.005  &     2.144  &   -9.521  &   -0.132  &   -9.017  &    0.447  &   -6.483  \\
10.0  &  10.00  &    2.981  &   -6.274  &     2.679  &   -4.460  &    1.876  &   -3.579  &    2.224  &   -2.607  \\
10.0  &  30.00  &    4.388  &    1.354  &     4.582  &    2.129  &    4.213  &    2.038  &    4.666  &    2.544  \\
11.0  &   0.01  &    5.525  & -100  &     4.591  & -100  &    4.801  & -100  &    4.328  & -100  \\
11.0  &   0.10  &    5.522  & -100  &     4.578  & -100  &    4.799  & -100  &    4.328  & -100  \\
11.0  &   0.20  &    5.524  & -100  &     4.558  & -100  &    4.801  & -100  &    4.328  & -100  \\
11.0  &   0.40  &    5.525  & -100  &     4.547  & -100  &    4.801  & -100  &    4.328  & -100  \\
11.0  &   0.70  &    5.529  & -100  &     4.542  & -100  &    4.801  & -100  &    4.328  & -100  \\
11.0  &   1.00  &    5.532  & -100  &     4.547  & -100  &    4.801  & -100  &    4.328  &  -96.327  \\
11.0  &   1.50  &    5.538  &  -84.221  &     4.601  &  -73.687  &    4.801  &  -74.596  &    4.328  &  -64.731  \\
11.0  &   2.00  &    5.543  &  -63.605  &     4.705  &  -55.599  &    4.801  &  -56.031  &    4.329  &  -48.785  \\
11.0  &   3.00  &    5.551  &  -42.822  &     4.921  &  -37.329  &    4.797  &  -37.263  &    4.334  &  -32.638  \\
11.0  &   5.00  &    5.562  &  -25.953  &     5.172  &  -22.468  &    4.774  &  -21.964  &    4.391  &  -19.431  \\
11.0  &  10.00  &    5.630  &  -12.792  &     5.455  &  -10.979  &    4.928  &  -10.095  &    5.057  &   -9.125  \\
11.0  &  30.00  &    6.228  &   -0.973  &     6.478  &   -0.198  &    6.093  &   -0.287  &    6.556  &    0.219  \\
\end{tabular}}
\end{table}
\begin{table}
\tbl{Same as Table 1 but for $^{54}$Ti, $^{55}$Ti and $^{55}$Ti}
{\scriptsize\begin{tabular}{|cc|cccccc|} $log\rho Y_{e}$ & $T_{9}$ &
\multicolumn{2}{|c|}{$^{54}$Ti}& \multicolumn{2}{|c|}{$^{55}$Ti} &
\multicolumn{2}{|c|}{$^{56}$Ti} \\
\cline{3-8} & &  \multicolumn{1}{c}{$\lambda_{\nu}$} &
\multicolumn{1}{|c|}{$\lambda_{\bar{\nu}}$} &
\multicolumn{1}{c}{$\lambda_{\nu}$} &
\multicolumn{1}{|c|}{$\lambda_{\bar{\nu}}$} &
\multicolumn{1}{c}{$\lambda_{\nu}$} &
\multicolumn{1}{|c|}{$\lambda_{\bar{\nu}}$} \\ \hline
 1.0  &   0.01  & -100  &   -0.174  & -100  &    0.296  & -100  &    1.531 \\
 1.0  &   0.10  & -100  &   -0.174  & -100  &    0.296  & -100  &    1.531 \\
 1.0  &   0.20  & -100  &   -0.174  & -100  &    0.296  & -100  &    1.531 \\
 1.0  &   0.40  & -100  &   -0.174  & -100  &    0.298  & -100  &    1.531 \\
 1.0  &   0.70  &  -89.296  &   -0.174  &  -89.048  &    0.351  & -100  &    1.531 \\
 1.0  &   1.00  &  -62.883  &   -0.174  &  -62.506  &    0.488  &  -74.004  &    1.531 \\
 1.0  &   1.50  &  -41.968  &   -0.174  &  -41.587  &    0.710  &  -49.274  &    1.531 \\
 1.0  &   2.00  &  -31.322  &   -0.174  &  -30.966  &    0.854  &  -36.724  &    1.531 \\
 1.0  &   3.00  &  -20.418  &   -0.169  &  -20.108  &    1.017  &  -23.904  &    1.531 \\
 1.0  &   5.00  &  -11.255  &   -0.081  &  -11.019  &    1.196  &  -13.197  &    1.542 \\
 1.0  &  10.00  &   -3.584  &    1.051  &   -3.470  &    1.542  &   -4.418  &    2.002 \\
 1.0  &  30.00  &    2.850  &    4.052  &    3.061  &    4.239  &    2.552  &    4.387 \\
 4.0  &   0.01  & -100  &   -0.175  & -100  &    0.296  & -100  &    1.531 \\
 4.0  &   0.10  & -100  &   -0.175  & -100  &    0.296  & -100  &    1.531 \\
 4.0  &   0.20  & -100  &   -0.175  & -100  &    0.296  & -100  &    1.531 \\
 4.0  &   0.40  & -100  &   -0.175  & -100  &    0.298  & -100  &    1.531 \\
 4.0  &   0.70  &  -86.929  &   -0.175  &  -86.680  &    0.351  & -100  &    1.531 \\
 4.0  &   1.00  &  -61.833  &   -0.175  &  -61.456  &    0.488  &  -72.954  &    1.531 \\
 4.0  &   1.50  &  -41.810  &   -0.175  &  -41.429  &    0.710  &  -49.116  &    1.531 \\
 4.0  &   2.00  &  -31.286  &   -0.174  &  -30.930  &    0.854  &  -36.689  &    1.531 \\
 4.0  &   3.00  &  -20.411  &   -0.169  &  -20.102  &    1.017  &  -23.898  &    1.531 \\
 4.0  &   5.00  &  -11.254  &   -0.081  &  -11.017  &    1.196  &  -13.195  &    1.542 \\
 4.0  &  10.00  &   -3.583  &    1.051  &   -3.469  &    1.542  &   -4.417  &    2.002 \\
 4.0  &  30.00  &    2.852  &    4.054  &    3.062  &    4.240  &    2.553  &    4.389 \\
 7.0  &   0.01  & -100  &   -0.289  & -100  &    0.240  & -100  &    1.498 \\
 7.0  &   0.10  & -100  &   -0.289  & -100  &    0.240  & -100  &    1.499 \\
 7.0  &   0.20  & -100  &   -0.289  & -100  &    0.240  & -100  &    1.499 \\
 7.0  &   0.40  & -100  &   -0.288  & -100  &    0.242  & -100  &    1.499 \\
 7.0  &   0.70  &  -80.628  &   -0.287  &  -80.379  &    0.301  &  -96.738  &    1.499 \\
 7.0  &   1.00  &  -56.835  &   -0.285  &  -56.458  &    0.448  &  -67.956  &    1.500 \\
 7.0  &   1.50  &  -38.028  &   -0.280  &  -37.648  &    0.680  &  -45.334  &    1.500 \\
 7.0  &   2.00  &  -28.465  &   -0.274  &  -28.109  &    0.828  &  -33.868  &    1.502 \\
 7.0  &   3.00  &  -18.701  &   -0.261  &  -18.392  &    0.996  &  -22.188  &    1.504 \\
 7.0  &   5.00  &  -10.551  &   -0.201  &  -10.314  &    1.179  &  -12.492  &    1.516 \\
 7.0  &  10.00  &   -3.484  &    0.999  &   -3.370  &    1.517  &   -4.318  &    1.978 \\
 7.0  &  30.00  &    2.855  &    4.050  &    3.066  &    4.237  &    2.557  &    4.386 \\
10.0  &   0.01  & -100  & -100  & -100  & -100  & -100  & -100 \\
10.0  &   0.10  &  -48.507  & -100  &  -51.341  & -100  & -100  & -100 \\
10.0  &   0.20  &  -27.864  & -100  &  -28.499  &  -88.596  &  -85.726  &  -94.766 \\
10.0  &   0.40  &  -16.932  &  -86.965  &  -16.466  &  -46.252  &  -45.055  &  -49.332 \\
10.0  &   0.70  &  -11.544  &  -50.916  &  -10.838  &  -27.654  &  -27.000  &  -29.408 \\
10.0  &   1.00  &   -8.808  &  -36.237  &   -8.361  &  -19.996  &  -19.407  &  -21.185 \\
10.0  &   1.50  &   -6.314  &  -24.554  &   -6.228  &  -13.836  &  -13.225  &  -14.510 \\
10.0  &   2.00  &   -4.890  &  -18.528  &   -5.028  &  -10.624  &   -9.953  &  -10.979 \\
10.0  &   3.00  &   -3.214  &  -12.247  &   -3.614  &   -7.232  &   -6.418  &   -7.185 \\
10.0  &   5.00  &   -1.171  &   -6.834  &   -1.259  &   -4.244  &   -3.020  &   -3.740 \\
10.0  &  10.00  &    1.589  &   -2.215  &    1.679  &   -1.564  &    0.758  &   -0.549 \\
10.0  &  30.00  &    4.367  &    2.541  &    4.577  &    2.727  &    4.068  &    2.917 \\
11.0  &   0.01  &    4.577  & -100  &    3.963  & -100  &    3.961  & -100 \\
11.0  &   0.10  &    4.580  & -100  &    3.963  & -100  &    3.959  & -100 \\
11.0  &   0.20  &    4.574  & -100  &    3.963  & -100  &    3.958  & -100 \\
11.0  &   0.40  &    4.577  & -100  &    3.963  & -100  &    3.959  & -100 \\
11.0  &   0.70  &    4.578  & -100  &    3.962  & -100  &    3.960  & -100 \\
11.0  &   1.00  &    4.578  & -100  &    3.961  &  -84.583  &    3.960  &  -85.621 \\
11.0  &   1.50  &    4.579  &  -67.600  &    3.957  &  -56.898  &    3.961  &  -57.458 \\
11.0  &   2.00  &    4.580  &  -50.818  &    3.955  &  -42.926  &    3.963  &  -43.194 \\
11.0  &   3.00  &    4.583  &  -33.783  &    3.955  &  -28.776  &    3.967  &  -28.667 \\
11.0  &   5.00  &    4.593  &  -19.774  &    3.972  &  -17.189  &    3.982  &  -16.635 \\
11.0  &  10.00  &    4.815  &   -8.724  &    4.556  &   -8.076  &    4.261  &   -7.014 \\
11.0  &  30.00  &    6.252  &    0.219  &    6.453  &    0.404  &    5.964  &    0.611 \\
\end{tabular}}
\end{table}
The summation over all initial and final states was carried out
until satisfactory convergence in the rate calculation was achieved.
The pn-QRPA theory allows a microscopic state-by-state calculation
of both sums present in Eqs. ~(\ref{nurate}) and ~(\ref{nubarrate}).
This feature of the pn-QRPA model greatly increases the reliability
of the calculated rates over other models in stellar matter where
there exists a finite probability of occupation of excited states.

The calculated neutrino and antineutrino energy loss rates due to
$^{49,51,52,53}$Ti are presented in Table 1 whereas Table 2 presents
the corresponding rates due to $^{54,55,56}$Ti. The calculated rates
are tabulated on an abbreviated density scale. The first column
gives log($\rho Y_{e}$) in units of $g cm^{-3}$, where $\rho$ is the
baryon density and $Y_{e}$ is the ratio of the electron number to
the baryon number. Stellar temperatures ($T_{9}$) are stated in
$10^{9} K$. $\lambda_{\bar{\nu}}$($\lambda_{\bar{\nu}}$) are the
neutrino(antineutrino) energy loss rates in units of  $MeV. s^{-1}$.
The calculated energy loss rates are tabulated in logarithmic (to
base 10) scale.  In the table, -100 means that the rate is smaller
than 10$^{-100} MeV. s^{-1}$. It can be seen from Table 1 that at
low densities and temperatures the antineutrino energy loss rates
due to $^{49,51,52,53}$Ti dominate by order of magnitudes and hence
more important for the collapse simulators. As T$_{9} [K] \sim 30$,
the neutrino energy loss rates try to catch up with the antineutrino
energy loss rates. At high stellar densities the story reverses with
neutrino energy loss rates assuming the role of the dominant
partner. At low densities the antineutrino energy loss rates have a
dominant contribution from the positron captures on $^{49}$Ti . As
temperature rises or density lowers (the degeneracy parameter is
negative for positrons), more and more high-energy positrons are
created leading in turn to higher positron capture rates and
consequently higher antineutrino energy loss rates. For the
remaining isotopes of titanium considered in this study the
antineutrino energy loss rates are dominated by the $\beta$-decay of
these isotopes, except when the stellar core attains high
temperature (T$_{9} [K] \sim 30$). The energy losses shown by
$^{51,52,53,54,55,56}$Ti follow a similar trend. At low densities
the antineutrino energy loss rates dominate and as the stellar core
stiffens to high densities, the neutrino energy loss rates become
more important for the collapse simulators. The neutrino and
antineutrino energy loss rates increases monotonically with
increasing stellar temperatures. From these tables it can be seen
that, e.g. at $\rho Y_{e} [gcm^{-3}] =10$ and T$_{9} [K] = 30$, the
neutrino and antineutrino energy loss per unit time per $^{51}$Ti
nucleus is 1161.4 $MeVs^{-1}$ and 4570.9 $MeVs^{-1}$, respectively.
The complete electronic version (ASCII files) of these rates may be
requested from the author.

\begin{figure}[th]
\centerline{\psfig{file=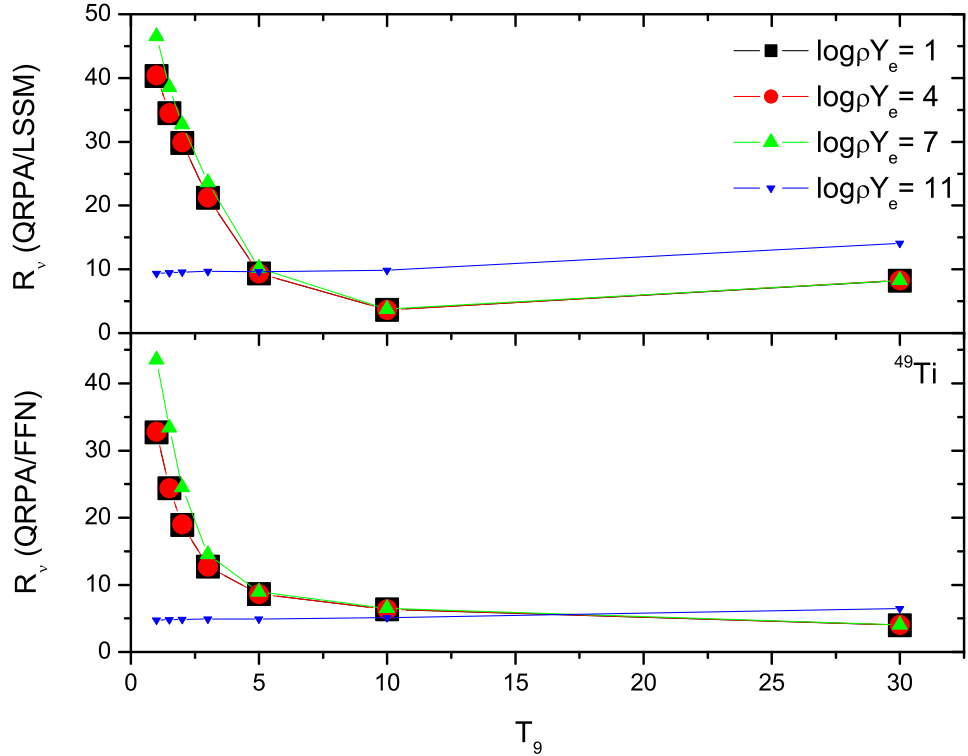, width=15cm}} \vspace*{8pt}
\caption{(Color online) Ratios of pn-QRPA neutrino energy loss rates
due to $^{49}$Ti to those calculated using LSSM (upper panel) and by
FFN (lower panel) as function of stellar temperatures and densities.
T$_{9}$ gives the stellar temperature in units of $10^{9}$ K. In the
legend, log $\rho Y_{e}$ gives the log to base 10 of stellar density
in units of $gcm^{-3}$.}\label{figure1}
\end{figure}
\begin{figure}[th]
\centerline{\psfig{file=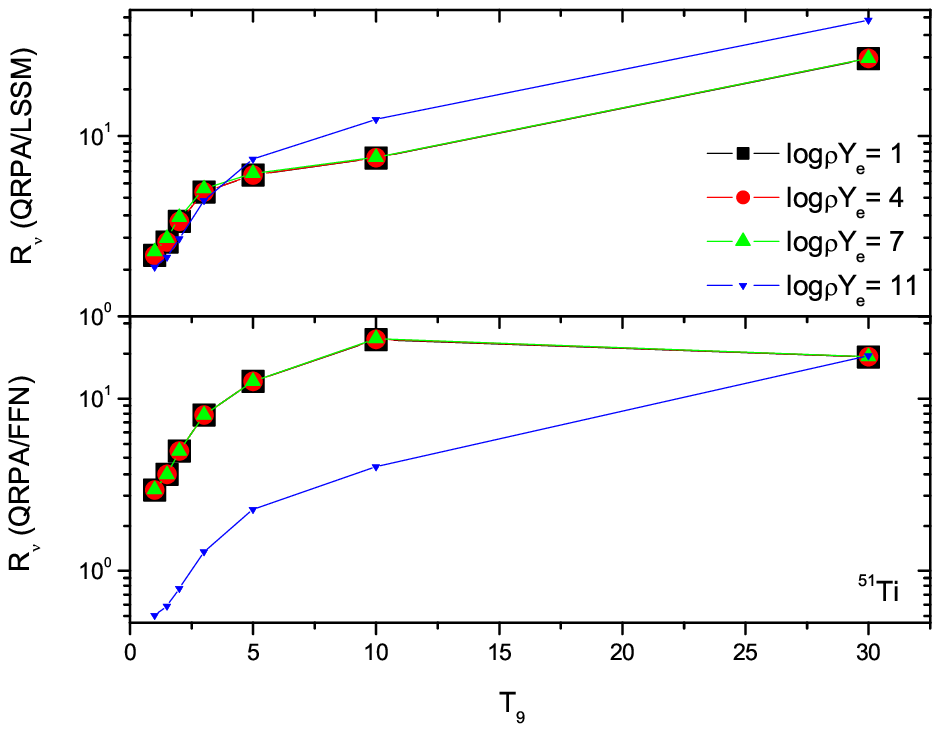, width=15cm}} \vspace*{8pt}
\caption{(Color online) Same as Figure 1 but for neutrino energy
loss rates due to $^{51}$Ti.}\label{figure2}
\end{figure}
\begin{figure}[th]
\centerline{\psfig{file=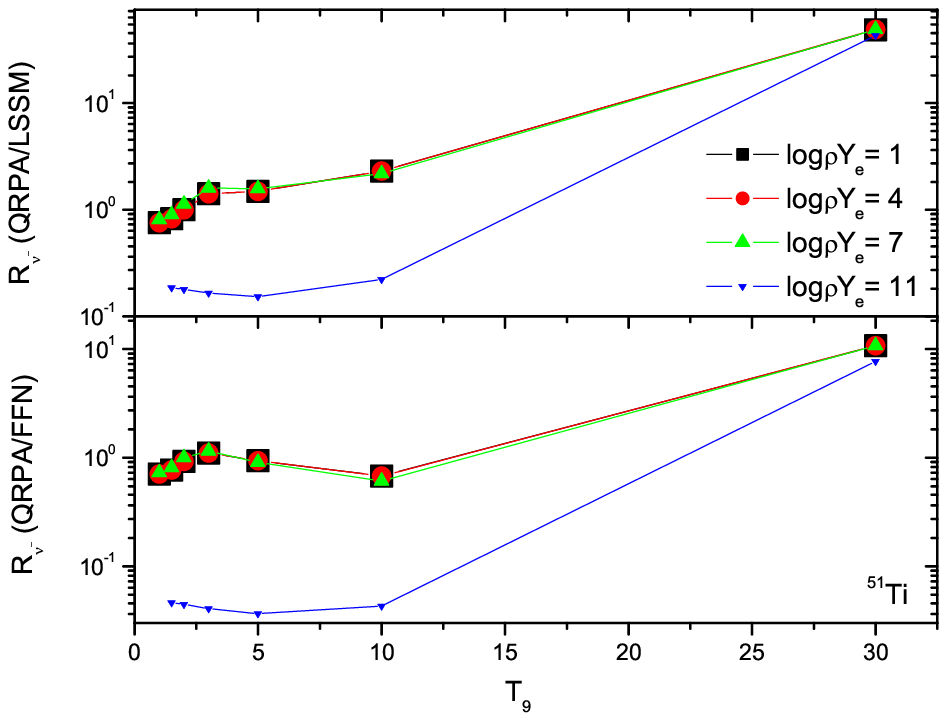, width=15cm}} \vspace*{8pt}
\caption{(Color online) Same as Figure 1 but for antineutrino energy
loss rates due to $^{51}$Ti.}\label{figure3}
\end{figure}
\begin{figure}[th]
\centerline{\psfig{file=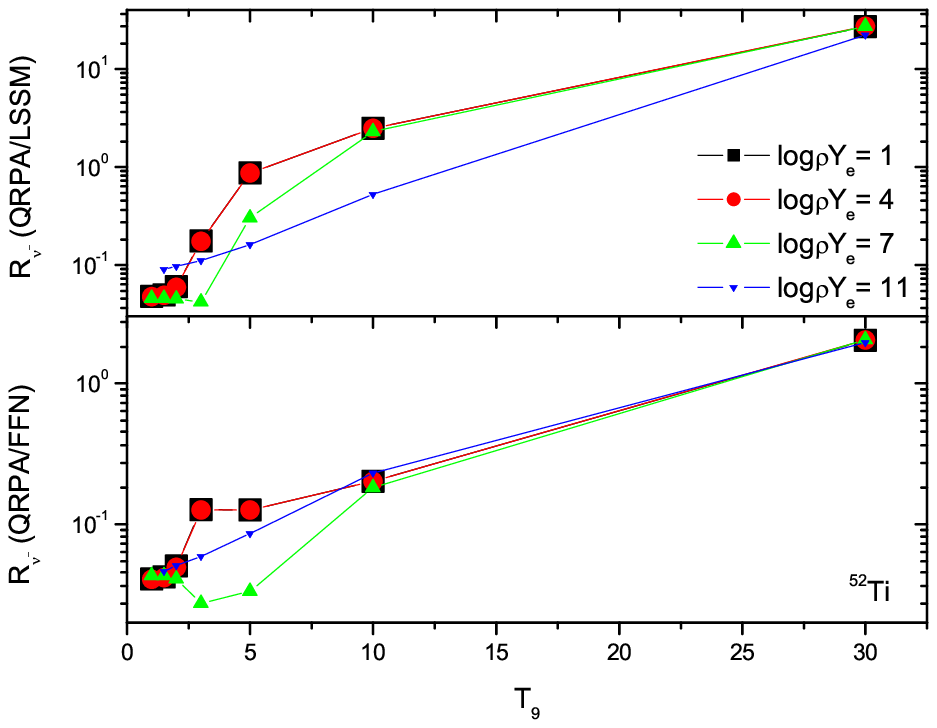, width=15cm}} \vspace*{8pt}
\caption{(Color online) Same as Figure 1 but for antineutrino energy
loss rates due to $^{52}$Ti.}\label{figure4}
\end{figure}
\begin{figure}[th]
\centerline{\psfig{file=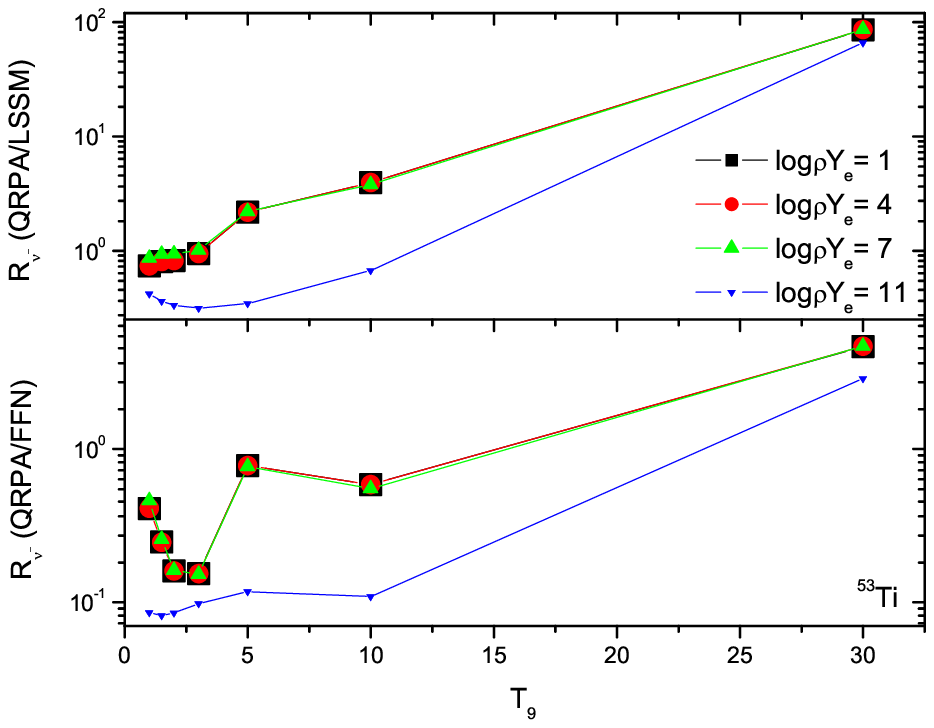, width=15cm}} \vspace*{8pt}
\caption{(Color online) Same as Figure 1 but for antineutrino energy
loss rates due to $^{53}$Ti.}\label{figure5}
\end{figure}
\begin{figure}[th]
\centerline{\psfig{file=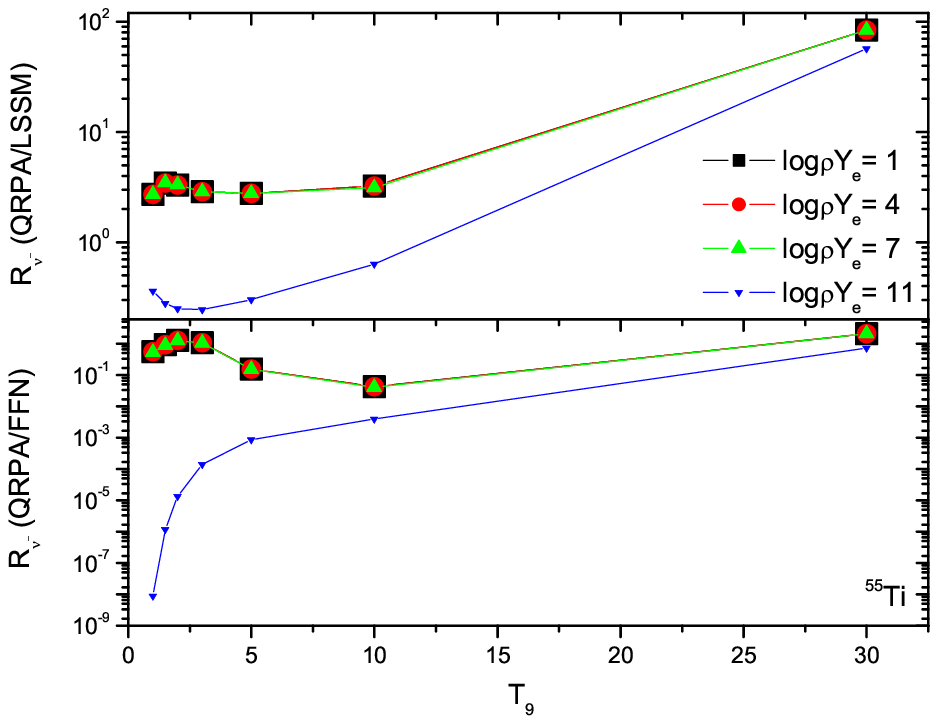, width=15cm}} \vspace*{8pt}
\caption{(Color online) Same as Figure 1 but for antineutrino energy
loss rates due to $^{55}$Ti.}\label{figure6}
\end{figure}
\begin{figure}[th]
\centerline{\psfig{file=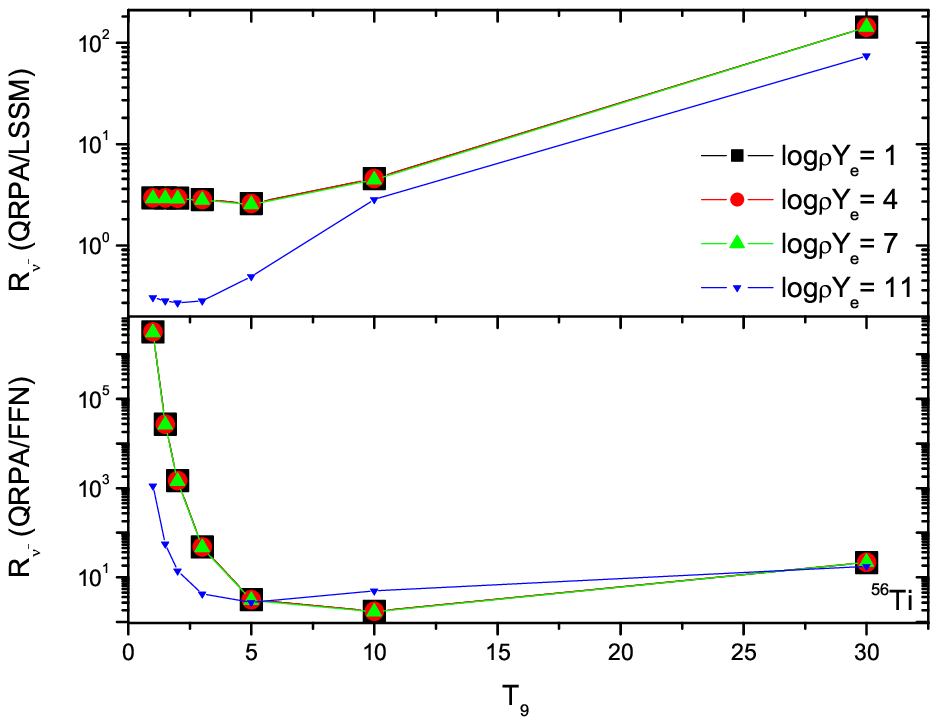, width=15cm}} \vspace*{8pt}
\caption{(Color online) Same as Figure 1 but for antineutrino energy
loss rates due to $^{56}$Ti.}\label{figure7}
\end{figure}
The calculation of neutrino and antineutrino energy loss rates was
also compared with previous calculations. For the sake of comparison
I took into consideration the pioneer calculation performed by FFN
\cite{Ful82} and those performed using the large-scale shell model
(LSSM) \cite{Lan00}. The FFN rates were used in many simulation
codes (e.g. KEPLER stellar evolution code \cite{Wea78}) while LSSM
rates were employed in recent simulation of presupernova evolution
of massive stars in the mass range 11-40 $M_{\odot}$ \cite{Heg01}.
Figure~\ref{figure1} depicts the comparison of neutrino energy loss
rates due to $^{49}$Ti with earlier calculations. Electron captures
on $^{49}$Ti during the core silicon burning phases of massive stars
are important due to the study performed by Aufderheide and
collaborators \cite{Auf94}. As such it is expected that the stellar
neutrinos produced as a result of electron capture on $^{49}$Ti may
contribute effectively in cooling the stellar core. The upper panel
displays the ratio of calculated rates to the LSSM rates,
$R_{\nu}(QRPA/LSSM)$, while the lower panel shows a similar
comparison with the FFN calculation, $R_{\nu}(QRPA/FFN)$. All graphs
are drawn at four selected values of stellar densities ($\rho Y_{e}
[gcm^{-3}] =10^{1}, 10^{4}, 10^{7}$ and  $10^{11}$). These values
correspond roughly to low, medium-low, medium-high and high stellar
densities, respectively. The selected values for temperature on the
abscissa are T$_{9} [K] = 1, 1.5, 2, 3, 5, 10$ and $30$. It can be
seen from Figure~\ref{figure1} that the pn-QRPA calculated neutrino
cooling rates are bigger than the corresponding LSSM rates by as
much as a factor of 50. At $\rho Y_{e} [gcm^{-3}] =10^{1}, 10^{4},
10^{7}$ the reported rates are bigger by at least a factor of 40 at
T$_{9} [K] = 1$. At T$_{9} [K] = 10$ the two rates are in reasonable
comparison (pn-QRPA rates are still bigger by a factor of 4). At
high densities the pn-QRPA rates are bigger roughly by an order of
magnitude. The pn-QRPA rates are also bigger than FFN rates (lower
panel) by as much as a factor of 44 at T$_{9} [K] = 7$ and $\rho
Y_{e} [gcm^{-3}] =10^{7}$. At high temperatures and densities the
rates are in reasonable comparison (within a factor of 5).

For the case of $^{51}$Ti the pn-QRPA and LSSM rates are in much
better agreement (Figure~\ref{figure2}). The upper panel shows that
till T$_{9} [K] \sim 5$ the rates are within a factor of five (with
pn-QRPA rates exceeding the LSSM rates). At higher temperatures the
pn-QRPA rates surpass the LSSM by as much as a factor of 44. A
similar comparison is shown in the lower panel of
Figure~\ref{figure2} where the pn-QRPA neutrino energy loss rates
are bigger by as much as a factor of 22 compared to FFN rates.

FFN did not calculate the neutrino energy loss rates due to
$^{52,53,54}$Ti whereas LSSM did not calculate the neutrino energy
loss rates due to $^{53,54}$Ti and as such a mutual comparison with
these previous calculations was not possible for the case of
$^{52,53,54}$Ti.

Next I move to the comparison of the antineutrino energy loss rates
with previous calculations. Here five cases, namely
$^{51,52,53,55,56}$Ti, were possible for mutual comparison with LSSM
and FFN rates (FFN did not calculate the antineutrino energy loss
rates due to $^{54}$Ti).

Figure~\ref{figure3} shows the comparison with LSSM and FFN rates
for the case of $^{51}$Ti. Here one sees that the LSSM and pn-QRPA
rates are in very good comparison at $\rho Y_{e} [gcm^{-3}] =10^{1},
10^{4}, 10^{7}$ (within a factor of 2) except at T$_{9} [K] = 30$
where the reported rates are bigger by around a factor of 50. At
high densities the LSSM rates are bigger roughly by a factor of 5.
At higher temperatures, T$_{9} [K] \sim 30$, the pn-QRPA rates are
again bigger by a factor of 43. The lower panel shows that the
reported antineutrino energy loss rates are in very good comparison
with the FFN rates at $\rho Y_{e} [gcm^{-3}] =10^{1}, 10^{4},
10^{7}$ with a perfect comparison at T$_{9} [K] = 3$. At higher
temperatures (T$_{9} [K] \sim 30$) the pn-QRPA rates are bigger by a
factor of 10. Only at high densities the FFN rates are bigger (by as
much as a factor of 25). However at T$_{9} [K] = 30$ the reported
rates are again bigger by a factor of 7.

The comparison of antineutrino energy loss rates for the case of
$^{52}$Ti is depicted in Figure~\ref{figure4}. Here one notes that
at low temperatures the LSSM rates are bigger by as much as a factor
of 20. The comparison improves as the stellar temperature increases.
Within the temperature range $5 \leq $ T$_{9} [K] \leq 10$ the
comparison is fairly good. The pn-QRPA rates keep enhancing as the
temperature of the stellar core increases. Finally at T$_{9} [K] =
30$ the pn-QRPA rates are bigger by as much as a factor of 27. At
higher temperatures excited state GT strength distributions are
required for the calculation of weak rates (parent excited states
have a finite probability of occupation). The LSSM employed the
so-called Brink's hypothesis in the electron capture direction and
back-resonances in the $\beta$-decay direction to approximate the
contributions from high-lying excited state GT strength
distributions. Brink's hypothesis states that GT strength
distribution on excited states is \textit{identical} to that from
ground state, shifted \textit{only} by the excitation energy of the
state. GT back resonances are the states reached by the strong GT
transitions in the inverse process (electron capture) built on
ground and excited states. On the other hand the pn-QRPA model
performs a microscopic calculation of the GT strength distributions
for \textit{all} parent excited states and provides a fairly
reliable estimate of the total stellar rates. A similar comparison
is observed against the FFN rates in the lower panel of
Figure~\ref{figure4}. The FFN rates are bigger by as much as a
factor of 25 at low temperatures. The comparison improves as the
stellar temperature increases and is very good at T$_{9} [K] = 30$
(within a factor of 2).

The pn-QRPA antineutrino energy loss rates due to $^{53}$Ti are in
good comparison with the corresponding LSSM rates (within a factor
of 3) as can be seen from Figure~\ref{figure5}. However at T$_{9}
[K] = 30$ the reported rates are bigger roughly by two orders of
magnitude for reasons mentioned before. FFN rates are bigger except
at T$_{9} [K] = 30$.

For the case of $^{55}$Ti, the pn-QRPA and LSSM rates are in good
comparison (see Figure~\ref{figure6}). Once again the reported rates
surpass the LSSM rates roughly by two orders of magnitude at T$_{9}
[K] = 30$. Looking at the lower panel one sees a staggering 8 orders
of magnitude bigger FFN rates at low temperatures and densities.
However for the same temperature and density domain the reported
rates are in good agreement with the corresponding LSSM rates
hinting towards the fact that FFN overestimated their antineutrino
energy loss rates. It is worth mentioning that these antineutrino
energy loss rates are very small numbers ($\sim 10^{-80}$) and can
change by orders of magnitude by a mere change of 0.5 MeV, or less,
in parent or daughter excitation energies and are more reflective of
the uncertainties in the calculation of energies. The comparison is
again good at T$_{9} [K] = 30$.

Finally I present the comparison of antineutrino energy loss rates
due to $^{56}$Ti with earlier calculations in Figure~\ref{figure7}.
The upper panel shows the fact that the pn-QRPA rates are in good
comparison with the LSSM rates (within a factor of 5) except at
T$_{9} [K] = 30$ where the reported rates are bigger than two orders
of magnitude. The FFN rates are smaller by around 6 orders of
magnitude at T$_{9} [K] = 1$. For the same physical conditions the
reported rates are in good comparison with the LSSM numbers again
hinting towards some problems in the FFN calculations. Unmeasured
matrix elements for allowed transitions were assigned an average
value of $log ft = $5 in FFN calculations. On the other hand these
transitions were calculated in a microscopic fashion using the
pn-QRPA theory (and the large scale shell model) and depict a more
realistic picture of the events taking place in stellar environment.

\section{Conclusions}

Isotopes of titanium are amongst the key iron-regime nuclei that
play a consequential role in the late phases of stellar evolution of
massive stars. The weak-interaction mediated reactions on these
nuclei, namely electron capture and $\beta$-decay, change the
lepton-to-baryon fraction ($Y_{e}$) during the late phases of
stellar evolution. The electron capture contributes in decreasing
$Y_{e}$ while the $\beta$-decay causes an increment in the $Y_{e}$
value. The temporal variation of $Y_{e}$ within the core of a
massive star has a pivotal role to play in the stellar evolution and
a fine-tuning of this parameter at various stages of presupernova
evolution is the key to generate an explosion. The neutrinos and
antineutrinos produced as a result of these weak interaction
reactions are transparent to the stellar matter at presupernova
densities and therefore assist in cooling the core to a lower
entropy state. A lower entropy environment can assist to achieve
higher densities for the ensuing collapse generating a stronger
bounce and in turn forming a more energetic shock wave. A search was
made from the literature to short list seven key titanium isotopes
in this respect. The pn-QRPA theory was employed to microscopically
calculate the neutrino and antineutrino energy loss rates due to
these seven titanium isotopes. The pn-QRPA model has two important
advantages as compared to other models. It can handle any
arbitrarily heavy system of nucleons since the calculation is
performed in a luxurious model space of up to 7 major oscillator
shells. Further it is the only available model that can calculate
\textit{all} excited state GT strength distributions in a
microscopic fashion which greatly increases its utility in stellar
calculations.

The neutrino and antineutrino energy loss rates were calculated on a
detailed density-temperature grid point and the ASCII files of the
rates can be requested from the author. The calculation was also
compared with the earlier pioneer calculations performed by FFN and
the recent microscopic large scale shell model calculation. The
reported neutrino energy loss rates are bigger by as much as a
factor of 44 as compared to LSSM rates at high stellar temperatures.
The corresponding antineutrino energy loss rates are bigger by more
than two orders of magnitude.

The enhanced pn-QRPA energy loss rates favor cooler cores with lower
entropies. This may affect the temperature, entropy and the
lepton-to-baryon ratio during the hydrostatic phases of stellar
evolution which becomes very important going into stellar collapse.
The core-collapse simulators are urged to test run the reported
stellar neutrino energy loss rates in core-collapse simulation codes
to check for some interesting outcome.

\end{document}